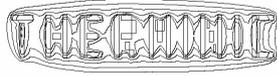

# DESIGN ISSUES OF A VARIABLE THERMAL RESISTANCE

*Székely V., Mezősi G*

Budapest University of Technology & Economics
Department of Electron Devices
szekely|mezosi@eet.bme.hu

**ABSTRACT**

A flat mounting unit with electronically variable thermal resistance has been constructed. The design is based on a Peltier cell and the appropriate control electronics and software. The device is devoted especially to the thermal characterization of packages, e.g. in dual cold plate arrangements.

The paper is dealing mainly with the dynamic behavior of the device and with the problems arising from the incertitude in Peltier model parameters. Finally experimental results are presented.

## 1. INTRODUCTION

Creation of boundary condition independent thermal package models is nowadays in the focus of the activity of many thermal engineers. This activity requires the measurement of packages under different boundary conditions. These boundary conditions appear as thermal resistance sets between the top/bottom/side of the package and the ambience. Usually a number of these resistance sets is fixed. These sets have to be applied during the measurements.

Changing the thermal boundary conditions requires the dismounting and mounting again the measurement setup, which is a tedious work. This fact raises the idea of a new tool: the mount with variable thermal resistance (VTR). By using such tools at the boundaries of the package, measurement for a number of different boundary conditions become possible in a single setup.

A few attempts to realize VTR structures can be found in the literature, but for very different applications. One solution is to apply thermally conductive fluid between two metal plates with interleaving fins. Change in the fluid quantity causes the change in the thermal resistance between the metal plates [1]. Another approach realizes ambient temperature dependent thermal resistance, using an array of bimorph cantilevers in a MEMS structure. The cantilevers deflect if the temperature raises. Since they have different length, they make contact with the opposite layer at different temperatures. This way the thermal resistance between the array and the opposite layer changes with the temperature [2].

Some years ago we have proposed a thermal mount with electronically variable thermal resistance, using Peltier cell [3]. In this earlier work the feasibility of such a structure has been demonstrated. Now we intend to realize this mount in a maturated form, suitable to the everyday use in the practice of package thermal qualification and modeling. The design of such a device raises a number of new questions and problems. The present paper is dealing with these problems and the possible solutions.

## 2. OPERATING PRINCIPLE

Let us briefly summarize the operating principle of the electronically variable thermal resistance. This mount is a sandwich structure, consisting of a Peltier cell, with a heat spreading plate in both sides (Fig.1). Temperature sensors are placed in both sides of the structure; the temperature data are forwarded to the control unit. This unit provides the driving current for the Peltier cell.

The Peltier current is controlled by the two temperatures $T_1$ and $T_2$ in such a way that the virtual thermal resistance "seen" on the top of the structure has to be the prescribed value. Due to the nonlinear behavior of the Peltier cell, this control is described by a nonlinear function, in order to achieve linear virtual thermal resistance.

The proper operation requires constant backside $T_2$ temperature. This is why mounting on a good cold plate is recommended.

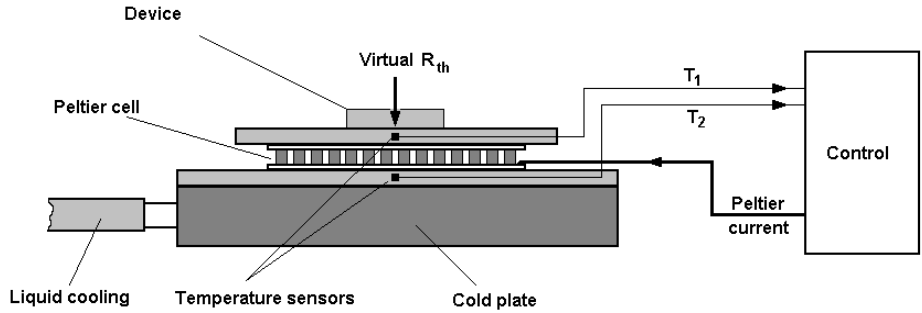

*Fig.1. The mount providing variable $R_{th}$*

## 3. MODELING OF THE PELTIER CELL

The theoretical investigation of the structure requires an appropriate model of the Peltier cell. In connection with our research work concerning the electro-thermal simulation algorithms we have developed a circuit model for the Peltier cell [3], [4]. This model is shown in Fig. 2. The model has the following three model parameters: $R$ is the electrical resistance in ohms, $R_{th}$ is the thermal resistance in K/W, $\alpha$ is the Peltier conversion constant in V/K. In Fig.2 $I$ denotes the electrical current, while $T_1$ and $T_2$ are the temperatures in the two sides of the cell, in Kelvin.

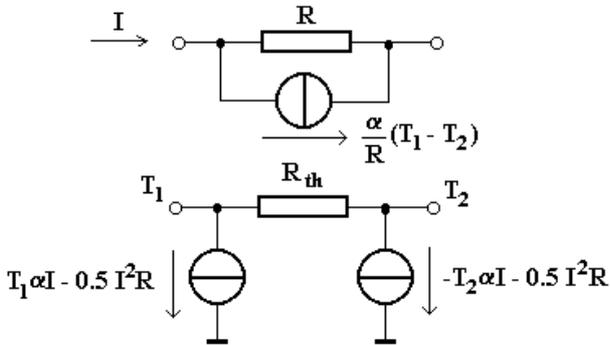

*Fig.2. Peltier cell model*

As an example, let us recall the parameters of a 40×40 mm surface, $I_{max}$ = 4A Peltier cell: $R_{th}$=2 K/W, $\alpha$=0.054 V/K, $R$=3.6Ω.

## 4. BASIC EQUATION OF THE OPERATION

The model of Fig. 2 has to be completed with the effect of the cold plate. That means $T_2$=constant (see Fig.3). The upper side has satisfy the equation that describes the linear virtual thermal resistance $R_{thv}$:

$$R_{thv} = \frac{T_1 - T_2}{P} \quad (1)$$

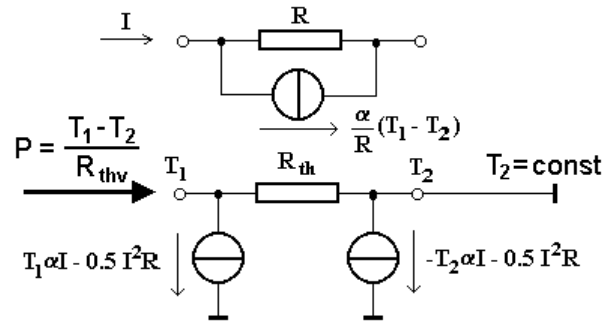

*Fig.3. Circuit model of the sandwich mount*

The following equation can be written to the $T_1$ node of the Fig.3:

$$0 = \frac{1}{2}I^2 R - T_1 \alpha I + \frac{T_1 - T_2}{R_{thv}} - \frac{T_1 - T_2}{R_{th}} \quad (2)$$

This is an equation of the second degree for the current $I$. Rearranging this equation leads to

$$I = \frac{T_1 \alpha \pm \sqrt{T_1^2 \alpha^2 - 2R(T_1 - T_2)(1/R_{thv} - 1/R_{th})}}{R} \quad (3)$$

This equation has to be used for the control of the current of the Peltier cell, in order to achieve a variable, linear virtual thermal resistance. Preliminary experiments show that the range of the realizable virtual thermal resistance extends to about two orders of magnitude, e.g. 0.1 – 10 K/W.

## 5. DYNAMIC BEHAVIOR

The circuit model of Fig. 3 has to be completed with two heat capacitances as it is shown in Fig.4. Since $T_2$=const the capacitance $C_{th2}$ has no effect. Unfortunately the heat capacitance $C_{th1}$ is slowing down the operation of the module, especially if we intend to realize relatively large virtual heat resistance. Under real circumstances $C_{th1}$ =10-20 Ws/K. If the virtual thermal resistance is 10 K/W, the time constant of the transient is 100-200 s and the settling time is much more high. (At the same time in case of $R_{thv}$=0.1 K/W the settling is only a few seconds.)

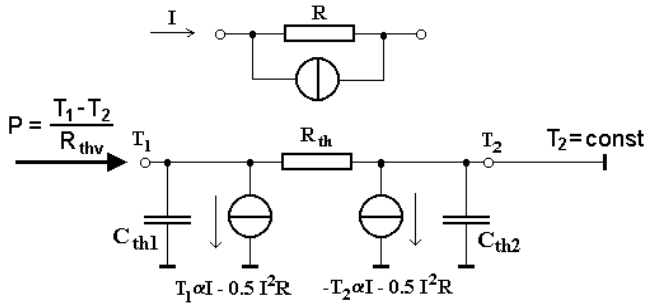

*Fig. 4. Dynamic model network*

In order to reduce the settling time the base equation of the control loop has to be completed with a dynamic part:

$$0 = \frac{1}{2}I^2 R - T_1 \alpha I + \frac{T_1 - T_2}{R_{thv}} - \frac{T_1 - T_2}{R_{th}} - m C_{th1} \frac{dT_1}{dt} \quad (4)$$

where *m* is a constant between 0 and 1. Using this equation, the dynamic behavior of the module is described by the following differential equation

$$\frac{dT_1}{dt} = \frac{1}{(1-m)C_{th1}} \left( P - \frac{T_1 - T_2}{R_{thv}} \right) \quad (5)$$

(Derivation of this equation is omitted here.) It is clearly visible that the solution of this equation is exponential but the time-constant of the transient is $(1-m)R_{thv}C_{th1}$ instead of $R_{thv}C_{th1}$. If we choose e.g. *m*=0.75, the settling will be four times faster. Approaching 1 with the value of *m*, however, holds the danger of oscillations in the control loop.

In order to use Eq. (4), of course, we have to measure (or calculate) the time derivative of the $T_1$ vs. time function.

## 6. AIM FOR THE ACCURATE MODEL PARAMETERS

The virtual thermal resistance is realized by the control based on the equation (3). The resulting $R_{thv}$ value is as accurate as this equation correct. Certainly, we know the Peltier model parameters with some incertitude. The same is true for the temperatures $T_1$ and $T_2$. An important question: in what extent this incertitude influences the $R_{thv}$ value?

Only one result of the detailed analysis is presented here. This is the effect of the incertitude in the $R_{th}$ value of the Peltier cell. The relative error of $R_{thv}$ can be approximately expressed as

$$\frac{\Delta R_{thv}}{R_{thv}} \cong \frac{R_{thv}}{R_{th}} \frac{\Delta R_{th}}{R_{th}} \quad (6)$$

This means that realizing 10 K/W virtual thermal resistance with the Peltier cell having $R_{th}$=2 K/W we have to known $R_{th}$ with 1 % accuracy to achieve 5 % accuracy in the value of virtual resistance.

## 7. MEASURE THE POWER FLUX

As an advantageous side effect the power flux streaming through the VTR module can be continuously displayed. The node equation for the node $T_1$ gives us:

$$P = -\frac{1}{2}I^2 R + T_1 \alpha I + \frac{T_1 - T_2}{R_{th}} + C_{th1} \frac{dT_1}{dt} \quad (7)$$

If we measure $T_1$, $T_2$ and $dT_1/dt$ we can calculate continuously the power flux across the module.

## 8. EXPERIMENTAL RESULTS

We have built the first version of the controllable thermal resistance unit. This unit consists of a 40×40 mm Peltier cell [5] and 80×80 mm Cu heat spreading plates. The total thickness of the unit is 12 mm. The two temperature sensors are forward biased pn junctions. The upper and the lower Cu plates are thermally insulated by using Teflon spacers. The photograph of the unit is shown in Fig. 5.

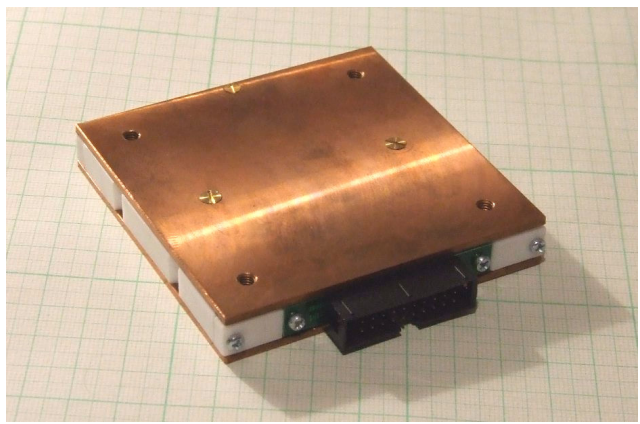

*Fig. 5. The variable thermal resistance (VTR) mount*

The control of the VTR unit is realized by software. The $T_1$ and $T_2$ temperature data are converted to digital ones and read in into the control program that runs on a PC. The program controls the current of the Peltier cell via an AD converter and a power voltage/current converter.

Let us see the results of a sequence of experiments! Three results are plotted in Figs. 6a,b and c. The Peltier current was controlled in order to achieve a.) 0.08 K/W, b.) 0.2 K/W and c.) 2 K/W thermal resistance. After the stabilization of the system a power pulse of 8 W has been forced by a transistor mounted on the unit (as shown in Fig. 1). The temperature rise divided by the 8W dissipation gives us the value of the realized virtual thermal resistance. The obtained resistance values are 0.06 K/W, 0.15 K/W and 1.65 K/W, respectively.

We can conclude that (i) the thermal resistance can be really controlled by using appropriate software control, (ii) the measured thermal resistance values differ from the expected ones with about 20-25 %, (iii) the settling time rises with increasing virtual thermal resistance. The error in the realized thermal resistance can be partly explained by the fact that some amount of the 8 W dissipation has been spread upwards instead of flowing across the VTR unit. Second possible source of the error is that the used Peltier parameters were probably not accurate enough.

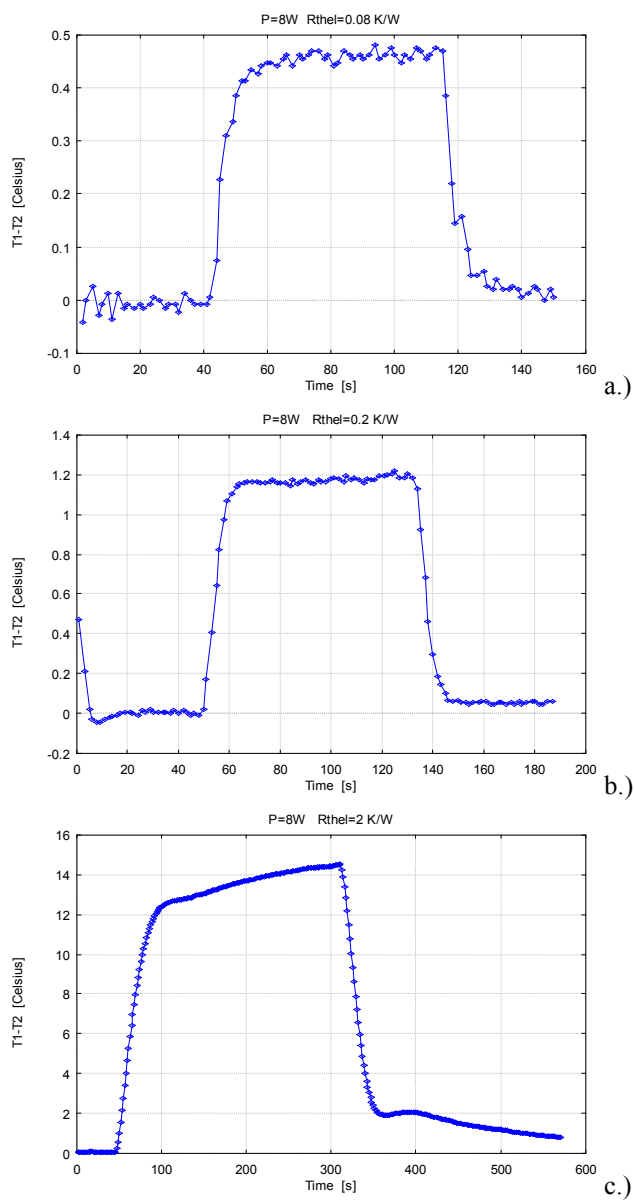

*Fig. 6 a,b,c. Temperature difference between the two sides of the unit if a power pulse of 8 W is applied*

Equation (7) provides us a method to measure the heat flux flowing across the VTR unit. The experiments show that this measurement of the power flux works quickly. Fig.7 shows the measured flux when 8 W dissipation is switched on/off in the device mounted on the unit (as shown in Fig. 1). The measured heat flux is about 7 W (according to the former assumption that not all of the 8 W dissipation has been crossed the VTR unit).

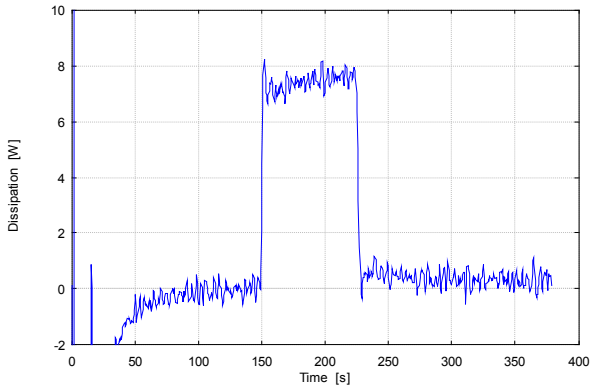

*Fig.7. Power flux streaming across the variable thermal resistance unit, measured using $T_1$, $T_2$ and $dT_1/dt$.*

## 9. CONCLUSIONS

The design of a variable thermal resistance (VTR) unit has been made. According to the former experiments, the measurements on the unit prove the feasibility of such a device. The experiments show that the thermal resistance can be varied between about 0.05 and 5 K/W for the actual design. The settling time is undesirably high above 2 K/W. In order to overcome this, during the design of an improved version the $C_{th1}$ thermal capacitance has to be reduced considerably.

Important progress has been achieved in the way to develop the practically applicable VTR device. The experiments help us to further improve the design. Further efforts are needed to extract more accurate parameters of the Peltier cells. Finally, the applications have to be tested/demonstrated during the subsequent research work.

## 10. ACKNOWLEDGMENTS

This work is supported by the "Gabor Baross" research and development innovation project of the Hungarian Government, KM-CSEKK-2005-000043.
The contribution of the Maxpert Ltd. in the realization of the experimental setup is very appreciated. Special thanks are expressed to S. Török for his effort in the design of control electronics.